\documentclass[11pt]{article}

\usepackage[active]{srcltx}
\usepackage{amsfonts}

\def\be{\begin{equation}}
\def\ee{\end{equation}}
\def\bea{\begin{eqnarray}}
\def\eea{\end{eqnarray}}
\def\noi{\noindent}
\def\nn{\nonumber}
\def\v{\tilde{X}_v}
\def\x{\tilde{X}_x}
\def\t{\tilde{X}_t}
\def\htt{\tilde{X}_{h_{00}}}
\def\htx{\tilde{X}_{h_{0x}}}
\def\hxx{\tilde{X}_{h_{xx}}}

\def\z{\tilde{X}_\phi}
\def \um {\frac{1}{2}}

\def\d{{\rm d}}

\begin{document}

\begin{center}
{\Large {\bf Symmetries of Non-Linear Systems: Group Approach to their Quantization}}
\end{center}

\bigskip
\bigskip

\centerline{V Aldaya$^{1}$, M Calixto$^{1,2}$, J Guerrero$^{1,3}$ and F F L\'opez-Ruiz$^{1}$}

\bigskip
\centerline{\it$^1$ Instituto de Astrof\'\i sica de Andaluc\'\i a (IAA-CSIC), }
\centerline{\it Apartado Postal 3004, 18080 Granada, Spain}
\centerline{\it $^2$ Departamento de Matem\'atica Aplicada, Facultad de Ciencias, Universidad de Granada, }
\centerline{\it Campus de Fuentenueva, 18071 Granada, Spain}
\centerline{\it $^3$ Departamento de Matem\'atica Aplicada, Facultad de Inform\'atica,  Universidad de  Murcia,}
\centerline{\it Campus de Espinardo,  30100 Murcia, Spain}

\bigskip

\centerline{valdaya@iaa.es $\quad$ calixto@ugr.es $\quad$ juguerre@um.es $\quad$ flopez@iaa.es}

\bigskip

\begin{center}

{\bf Abstract}
\end{center}

\small
\setlength{\baselineskip}{12pt}

\begin{list}{}{\setlength{\leftmargin}{3pc}\setlength{\rightmargin}{3pc}}
\item
We report briefly on an approach to quantum theory entirely based on
symmetry grounds which improves Geometric Quantization in some respects
and provides an alternative to the canonical framework. 
The present scheme, being typically non-perturbative,  
is primarily intended for non-linear systems,
although needless to say that finding the basic symmetry associated with a
given (quantum) physical problem is in general a difficult task, which many
times nearly emulates the complexity of finding the actual (classical)
solutions. Apart from some interesting examples related to the
electromagnetic and gravitational particle interactions, where an
algebraic version of the equivalence principle naturally arises, we
attempt to the quantum description of non-linear sigma models. In particular, we present 
the actual quantization of the partial-trace non-linear $SU(2)$ sigma model as a
representative case of non-linear quantum field theory.

\end{list}

\normalsize

\noi PACS: 11.30.-j, 11.30.Na, 03.65.Fd, 03.65.-w, 03.70.+k, 02.20.Tw, 04.20.Fy
\setlength{\baselineskip}{14pt}

\section{Introduction}

In this brief report we attempt to illustrate the features of a group-theoretical 
approach to the quantum description of fundamental physical systems, which is being developed 
over the last decades \cite{23}. The underlying motivation for pushing ahead the present 
Group Approach to Quantization (GAQ) is twofold. On the one hand we face a basic argument 
of beauty; we attempt to contribute to the big effort that had been devoted to place Quantum 
Mechanics in a similar geometrical status to that of Classical Mechanics or, 
even, General Relativity, in that which was known as Geometric Quantization 
(GQ) \cite{GQ1,GQ2,GQ3,GQ4}. On the other, there exists more practical reasons 
supported by a great variety of physical examples where the old (canonical) quantization 
finds serious difficulties in describing them properly. In fact,  the limitations 
of Canonical Quantization were soon stated neatly 
through the so-called ``no-go'' theorems \cite{no-go} (see also \cite{no-go2}).

The main ingredient in GAQ is the group structure taken to the ultimate 
consequences, that is to say, symmetry is intended to contribute to 
Physics as a building block rather than a practical tool for finding 
additional solutions to partially solved (symmetrical) problems. Even 
more, this approach attempts to describe a quantum physical system from 
the group manifold itself and its canonical structures, aiming at reducing 
the problem of establishing the physical postulates to that of choosing 
specific groups. In addition, it should be considered as a method for describing directly
the quantum dynamics since the intermediate step of solving the classical equation 
of motion is not required. In fact, the quantum nature of a given system can be 
associated with the actual (compact) topology of (part of) the addressing symmetry 
group, whereas the classical limit is obtained by simply taking a local version 
(in the sense of taking a local chart) of the symmetry (``opening'' the multiplicative 
$U(1)$ central subgroup to the additive real line $\mathbb R$). 

From a technical point of view, the present method also represents  
significant advantages. In particular, the biggest obstruction found by 
Geometric Quantization in dealing with non-linear systems, that of 
achieving the complete reduction of the geometric representation 
(polarization), can now be much better addressed on the grounds of the 
algebraic group structure. This is mainly due to the existence of two 
mutually commuting (left- and right-) actions, so that the infinitesimal generator 
of one of them can be used to construct the Poisson (classical) algebra 
representation (pre-quantization in the sense of Geometric Quantization), 
whereas the other can be employed to reduce completely the representation (true 
quantization). 

Although the requirement of the additional structure of Lie group might be seen as a 
drawback, it should be remarked that after all, the Lie algebra structure  is one of 
the few bricks shared by all quantization methods, which look for unitary and 
irreducible representations of a given Lie (Poisson) algebra somehow characterizing 
a physical system.

This paper is organized as follows. In Sec. 2 we motivate the central 
extensions of classical symmetry groups with the example of the Galilei 
group as well as the extension of classical phase space with an extra 
variable. In Sec. 3 the fundamentals of the Group Approach to Quantization 
scheme are presented. Sec. 4 is devoted to illustrating the way in which 
GAQ describes physical systems bearing a finite number of degrees of 
freedom. We start with the example of the free particle and then proceed 
to introduce interactions through some sort of revisited Minimal Coupling 
Principle. In particular, the particle moving in an electromagnetic field, 
as well as the geodesic motion in a gravitational field are analyzed. In 
the last case, a very simple algebraic version of the Equivalence 
Principle naturally arises. In Sec. 5 we present examples of infinite-dimensional
systems. After studying the example of the Klein-Gordon field we end up with
the case the Non-Linear Sigma Model (NLSM) as an example of a genuine non-linear
field.    

We wish to mention that the present GAQ method has been applied to numerous physical
systems that can not be reported here. Among them, we refer the reader to 
Refs. \cite{relatoscil,Klauder,Pepe1,radvacio,halleff,massorigin}.
 
\section{The role of central extensions of ``classical" symmetries}

Let us consider the symmetry of the Lagrangian of the free
particle in $1+1$ dimensions:\\
\[ {\cal L}=\frac{1}{2}m\stackrel{.}{x}^2\]

\noi Under the classical Galilean transformations
\be x'= x+A+Vt,\;\; t'= t+B\label{gali}, \ee The Lagrangian moves to
${\cal L}'=\frac{1}{2}m(\stackrel{.}{x}+V)^2={\cal L}\;
+\;\frac{\!\!d}{dt}
   (\frac{1}{2}mV^2t+mVx)$. That is, ${\cal L}$ is not strictly invariant, but {\it semi-invariant}, due to the presence of the total
derivative.

In infinitesimal terms something similar happens. Taking the Lie
derivative of  {${\cal L}$} with respect to the generators of the group
results in:
\bea
X_B&=&\frac{\partial}{\partial t}\qquad \quad \Rightarrow\quad X_B{\cal L}=0\nn\\
X_A&=&\frac{\partial}{\partial x}\qquad \quad \Rightarrow\quad X_A{\cal L}=0  \nn\\
X_V&=&t\frac{\partial}{\partial x}\; + \frac{\partial}{\partial \dot x}\Rightarrow \quad X_V{\cal L}=
{\frac{\!\!d}{dt}(mx)\neq 0}\nn \eea
\noi The same is also valid for the  {\it Poincar\'e-Cartan form}:
\[
 {\Theta_{PC}=pdx-Hdt}=\frac{\partial{\cal L}}{\partial\stackrel{.}{x}}dx
-(\stackrel{.}{x}p-{\cal L})dt=\frac{\partial{\cal
L}}{\partial\stackrel{.}{x}}(dx-\stackrel{.}{x}dt)+{\cal L}dt \]
whose Lie derivative is:  \be L_{X_B}\Theta_{PC}=0,\;\;
L_{X_A}\Theta_{PC}=0,\;\; L_{X_V}\Theta_{PC}= {d(mx)\neq 0}.\ee

The quantum free particle suffers from the same ``pathology" although it
manifests in a different manner. Let us apply the Galilean transformations
(\ref{gali}) to the Schr\"odinger equation. We get:\\
\[ i\hbar\frac{\partial\Psi}{\partial t}=-\frac{\hbar^2}{2m}\nabla^2\Psi\;\;\rightarrow\;\;
i\hbar\frac{\partial\Psi}{\partial t'}=-\frac{\hbar^2}{2m}{\nabla'}^2\Psi
 {-i\hbar V\frac{\partial\Psi}{\partial x'}}. \]
The extra term can be compensated if we also transform the wave function
by means of a non-trivial {phase}:
\be \Psi\to \Psi'= {e^{i\frac{m}{\hbar}(Vx+\frac{1}{2}V^2t)}}\Psi.
\label{psilaw} \ee
Then, we recover the original (fully primed) Schr\"odinger equation
${i\hbar\frac{\partial\Psi'}{\partial t'}=-\frac{\hbar^2}{2m}
{\nabla'}^2\Psi'}$.
Joining together the Galilean transformations (\ref{gali}) and the phase
transformation (\ref{psilaw}) we obtain a group of {\it strict} symmetry
whose group law is:
\bea
B''&=&B'+B \nn\\
A''&=&A'+A+V'B \nn\\
V''&=&V'+V \label{Galileo}\\
\zeta''&=&\zeta'\zeta
{e^{i\frac{m}{\hbar}[A'V+B(V'V+\frac{1}{2}{V'}^2)]}},\nn \eea
\noi where the last line has the general form $\zeta''=\zeta'\zeta
e^{i\frac{m}{\hbar}\xi(g',g)}$, with $\zeta\equiv e^{i\phi}\in U(1)$ and
the function $\xi$ being that which is customarily named $2$-cocycle on
the Galilei group, characterized by the mass $m$ \cite{Bargmann,Levy,Sudarshan}.
A constant $\hbar$ with the dimensions of an action has to be introduced to keep
the exponent dimensionless. 

The infinitesimal version of the group law (\ref{Galileo}) is expressed by means of the extended
Lie algebra commutators:
\be \left[\tilde{X}_B,\;\tilde{X}_A\right]=0,\;\;
\left[\tilde{X}_B,\;\tilde{X}_V\right]=\tilde{X}_A,\;\;
\left[\tilde{X}_A,\;\tilde{X}_V\right]=-m\tilde{X}_\phi. \ee
between the extended generators:
\bea
\tilde{X}_B&=&\frac{\partial}{\partial t}     \nn\\
\tilde{X}_A&=&\frac{\partial}{\partial x}      \nn\\
\tilde{X}_V&=&t\frac{\partial}{\partial x} + \frac{\partial}{\partial \dot x}-mxi\zeta\frac{\!\!\partial}{\partial\zeta} \nn\\
\tilde{X}_\phi&=&i\zeta\frac{\!\!\partial}{\partial\zeta}\,.\nn
\eea 

One of the relevant points concerning both the {\it strict invariance}
and, consequently, the {\it centrally extended symmetry} is that the
corresponding {\it extended Lie algebra} now properly represents  the
Poisson algebra generated by $\langle H\equiv\frac{P^2}{2m},\,K\equiv x-\frac{p}{m}t,\,P\equiv p,\,1 \rangle$ when
acting as ordinary derivations on complex functions, provided that we
impose that the new generator $\tilde{X}_\phi$ acts on $\Psi$ as
${\tilde{X}_\phi\Psi=i\Psi}$, or, in finite terms, $\Psi(\zeta g)=\zeta \Psi(g)$. 
Notice that the unextended algebra, with the
commutator $[X_A,\;X_V]=0$, is not an isomorphic image of the
corresponding Noether invariants $H,P,K$ algebra.

There is yet another remarkable advantage of requiring the strict symmetry
of a given arbitrary classical system. In fact, such a symmetry can only
be realized faithfully if we extend the classical phase space $M$
parameterized by $K, P$ (or solution manifold) by an extra variable, to be
identified with $\phi$ or $\zeta=e^{i\phi}$. In the compact ($U(1)$) case,
that is, the choice of $\zeta$, we thus arrive at the notion of a {\it
quantum manifold} $Q$ \cite{GQ1,GQ4}. In this manifold, locally parameterized
by $K,P,\zeta\equiv e^{i\phi}$, an extended Liouville form ($\Theta_{PC}$ 
defines the Liouville form $\vartheta$ on the solution manifold except for a 
total differential)
\[\Theta=\vartheta+\frac{d\zeta}{i\zeta} \;\;\; (\hbox{or}\;\; PdK+d\phi) \]
\noi substitutes successfully the ordinary one in the search for an
invertible duality between Hamiltonian functions and Hamiltonian vector
fields. In fact, the Hamiltonian correspondence
\[f\;\mapsto\;X_f \;\;\hbox{such that}\;\;i_{X_f}d\vartheta=-df\]

\noi has the real numbers $\mathbb R$ as kernel. However the
correspondence
\[f\;\mapsto\;\tilde{X}_f \;\;\hbox{such that}\;\;i_{\tilde{X}_f}d\Theta=-df\,,
i_{\tilde{X}_f}\Theta=f\]
\noi has unique solution. 

This is, so to speak, the starting point for GQ, where the 
pair (symplectic manifold) $(M,\omega\equiv d\vartheta)$ is replaced with the pair $(Q,\Theta)$
as a $U(1)$-principal bundle with connection (quantum manifold) under the requirement that
the curvature of $\Theta$ defines on $M$ the symplectic form $\omega$ with integer 
co-homology class (tantamount to say that the integration of $\vartheta$ on closed curves results in  
an integer; this is a modern, geometric version of the Bohr-Sommerfeld rules \cite{GQ4}). 
The association $f\;\mapsto\;\hat{f}\approx \tilde{X}_f$ defines the pre-quantum operators, as 
derivations on complex $U(1)$-functions on $Q$, which realize a unitary representation of the 
Poisson bracket although non-irreducible. The true quantization, that is to say, 
the irreducibility, is intended to be achieved after the polarization condition is imposed (see
Refs. \cite{GQ1,GQ4} and the analogous condition in next section).

It should be mentioned that the possibility exists of extending the classical
phase space by the real line and the classical group by the non-compact additive
group $\mathbb R$. In that case the constant $\hbar$ is no longer needed and the
resulting theory describes the classical limit in a \textit{global} version of the 
Hamilton-Jacobi formulation (see Ref. \cite{23}).

\section{Group Approach to Quantization}
The essential idea underlying a group-theoretical framework for
quantization consists in selecting a given subalgebra $\tilde{\cal G}$ of
the classical Poisson algebra including $ \langle
H,\,p_i,\,x^j,\,1\rangle$ and finding its {\it unitary irreducible
representations} (unirreps), which constitute the possible {\it
quantizations}. Although the actual procedure for finding unirreps might
not be what really matters from the physical point of view we proceed
along a well-defined algorithm, the group approach to quantization or GAQ
for brief, to obtain them for any Lie group.

All the ingredients of GAQ are canonical structures defined on Lie groups
and the very basic ones consist in the {\it two} mutually {\it commuting}
copies of the Lie algebra $\tilde{\cal G}$ of a group $\tilde{G}$ of {\it
strict symmetry} (of a given physical system), that is, the set of left-
and right-invariant vector fields:

\[{\cal X}^L(\tilde{G})\approx\tilde{\cal G}\approx{\cal
X}^R(\tilde{G})\]
\noi  in such a way that one copy, let us say ${\cal X}^R(\tilde{G})$,
plays the role of {\it pre-Quantum Operators} acting (by usual derivation)
on complex (wave) functions on $\tilde{G}$, whereas the other, ${\cal
X}^L(\tilde{G})$, is used to {\it reduce} the representation in a manner
{\it compatible} with the action of the operators, thus providing the {\it
true quantization}.

In fact, from the group law $g''=g'*g$ of any group $\tilde{G}$, we can
read two different left- and right-actions:
\be g''=g'*g\equiv L_{g'}g,\;\;\;  g''=g'*g\equiv
R_{g}g'.\label{leftrightact}\ee
\noi Both actions commute and so do their respective generators $\tilde{X}^R_a$ and
$\tilde{X}^L_b$, i.e.
$[\tilde{X}^L_a,\;\tilde{X}^R_b]=0\;\;\forall a,b$.


Another manifestation of the commutation between left an right
translations corresponds to the invariance of the left-invariant canonical
1-forms, $\{{\theta^L}^a\}$ (dual to $\{\tilde{X}^L_b\}$, i.e.
${\theta^L}^a(\tilde{X}^L_b)=\delta^a_b$) with respect to the
right-invariant vector fields, that is: $L_{\tilde{X}^R_a}{\theta^L}^b=0$
and the other way around ($L\leftrightarrow R$). In particular, we dispose
of a natural {\it invariant volume} $\omega$ on the group manifold since
we have:
\begin{equation}
L_{\tilde{X}^R_a}({\theta^L}^b\wedge{\theta^L}^c\wedge{\theta^L}^d...)\equiv
L_{\tilde{X}^R_a}\omega=0\,.
\label{omega}
\end{equation}

We should then be able to recover all {\it physical ingredients} of
quantum systems out of {\it algebraic structures}. In particular, the
Poincar\'e-Cartan form $\Theta_{PC}$ and the phase space itself
$M\equiv(x^i,p_j)$ should be regained from a group of {\it strict symmetry
$\tilde{G}$}. In fact, in the special case of a Lie group which bears a
central extension with structure group $U(1)$ parameterized by
$\zeta\in C$ such that $|\zeta|^2=1$, as we are in fact considering, the
group manifold $\tilde{G}$ itself can be endowed with the structure of a
{\it principal bundle} with an {\it invariant connection}, thus
generalizing the notion of {\it quantum manifold}.

More precisely, the $U(1)$-component of the left-invariant canonical form
(dual to the {\it vertical} generator $\tilde{X}^L_\zeta$, i.e.
$\theta^{L(\zeta)}(\tilde{X}^L_\zeta)=1$) will be named {\it quantization
form} $\Theta\equiv{\theta^L}^{(\zeta)}$ and generalizes the 
Poincar\'e-Cartan form $\Theta_{PC}$ of Classical Mechanics. The 
quantization form remains {\it strictly invariant} under the group 
$\tilde{G}$ in the sense that 
\[L_{\tilde{X}^R_a}\Theta=0 \ \ \forall a\]

\noi whereas $\Theta_{PC}$ is, in general, only {\it semi-invariant}, that
is to say, it is invariant except for a total differential.

It should be stressed that the construction of a true quantum manifold
in the sense of Geometric Quantization \cite{GQ1,GQ2} can
be achieved by taking in the pair $\{\tilde{G},\,\Theta\}$ the quotient by
the action of the subgroup generated by those left-invariant vector fields in the kernel of $\Theta$ and
$d\Theta$, that which is called in mathematical terms {\it characteristic
module} of the $1$-form $\Theta$,

\[ {\cal C}_\Theta\;\equiv\;\{\tilde X^L\;/\; i_{\tilde X^L}d\Theta=0=i_{\tilde X^L}\Theta\}.\]
A further quotient by structure subgroup $U(1)$ 
provides the {\it classical solution Manifold} $M$ or classical {\it phase space}. Even
more, the vector fields in ${\cal C}_\Theta$ constitute the (generalized) {\it classical equations of motion}.

On the other hand, the right-invariant vector fields are used to provide
classical functions on the phase space. In fact, the functions
\be 
F_a\;\;\equiv\;\; i_{\tilde{X}^R_a}\Theta
\label{Noether}
\ee
\noi are stable under the action of the left-invariant vector fields in
the characteristic module of $\Theta$, the equations of motion,
\[ L_{\tilde X^L}F_a=0\;\;\forall \tilde X^L\in {\cal C}_\Theta\]

\noi and then constitute the {\it Noether invariants}.

As a consequence of the central extension structure in $\tilde{G}$ the
Noether invariants (and the corresponding group parameters) are classified
in basic (symplectic or dynamical) and non-basic (non-symplectic or kinematic) depending
on whether or not the corresponding generators produce the central
generator by commutation with some other. Basic parameters (Noether
invariants) are paired (and independent). Non-basic Noether invariants
(like energy or angular momenta) can be written in terms of the basic ones
(positions and momenta).

As far as the quantum theory is concerned, the above-mentioned quotient by
the classical equations of motion is really not needed. We consider the
space of complex functions $\Psi$ on the whole group $\tilde{G}$ and
restrict them to only $U(1)$-functions, that is, those which are
homogeneous of degree $1$ on the argument $\zeta\equiv e^{i\phi}\in U(1)$.
Wave functions thus satisfy the $U(1)$-function condition
\be {\tilde{X}^L_\phi\Psi=i\Psi}.\label{u1function}\ee
On these functions the right-invariant vector fields act as {\it
pre-quantum operators} by ordinary derivation. They are, in fact,
Hermitian operators with respect to the scalar product with measure given by the 
invariant volume $\omega$ defined above (\ref{omega}).  However, this action is 
not a proper quantization of the Poisson algebra of the Noether invariants 
(associated with the symplectic structure given by $d\Theta$) since there is a set 
of non-trivial operators commuting with this representation. In fact, all the
left-invariant vector fields do commute with the right-invariant ones,
i.e. the pre-quantum operators,and, therefore, the representation is not irreducible. 
According to Schur's Lemma those operators must be trivialized. To this end we define a 
polarization subalgebra as follows:

\noi {\it A polarization} ${\cal P}$  {\it is a maximal left subalgebra
containing the characteristic subalgebra ${\cal G}_\Theta$ and excluding
the central generator}.

\noi The role of a polarization is that of {\it reducing} the
representation which then constitutes a true {\it quantization}. To this
end we impose on wave functions the polarization condition:
\[ \tilde{X}^L_b\Psi=0\, \ \ \forall \tilde{X}^L_b\in {\cal P}\]
In finite terms the polarization condition is expressed by the invariance of the wave 
functions under the finite action of the Polarization Subgroup $G_P$ acting from the right, that is:
\be
\Psi(g'g_P)=\Psi(g')\ \ \ \forall g_P\in G_P\,. \label{PolarizacionF}
\ee

To be intuitive, a polarization is made of half the left-invariant vector
fields associated with basic (independent) variables of the solution
manifold in addition to those associated with non-symplectic parameters as time
or rotational angles. We should remark that the classification
above-mentioned of the Noether invariants in basic and non-basic also
applies to the quantum operators so that the latter ones are written in
terms of the formers.

As an additional comment regarding polarization conditions, it must be
stressed that when expressed as quantum equations, they contain, in
particular, the evolution equation properly, that is, the
Schrodinger(-like) equation. In this respect these polarizations (and the
GAQ method itself) depart from those in Geometric Quantization, which are
imposed only after having taken the quotient by the classical evolution
explicitly, that which means having solved the classical equations. 
Another respect on which GAQ departs from GQ is in that the entire 
enveloping algebra (both left and right ones) can be used to construct higher-order 
Polarizations and higher-order operators.

The integration volume $\omega$ can be restricted to the Hilbert space of
polarized wave functions ${\cal H}$ by means of a canonical procedure a
bit technical for the scope of the present report. We refer the reader to
Ref.\cite{medida}.

Before ending this section let us mention that the existence of a
polarization containing the entire characteristic subalgebra (usually
referred to as full polarization) is not guaranteed in general and we then
can resort to the left enveloping algebra to complete the polarization in
the same way that any operator in the right enveloping algebra can be
properly realized as a quantum operator (see Ref. \cite{Higherpola}).
Higher-order polarizations are used by strict necessity, when no full
polarization can be found (in this case the system is {\it anomalous} in
the standard physical sense \cite{Jakiw}), or simply by pure convenience of 
realizing the quantization in a particular ``representation" adapted to 
given variables.

\section{Quantum Mechanics (examples with a finite number of degrees of freedom)}
In this section some examples of {\it symmetry groups $\leftrightarrow$ physical systems}
correspondence involving a finite number of dynamical variables are reported in the
simplest manner, showing the way GAQ can be used in practice. Formal developments
and subtleties are left for a further reading of the references.

\subsection{The Free Galilean Particle}
We shall adopt the notation $B\equiv t,\;A\equiv x,\;V\equiv v$ ($p\equiv
mv$) for the parameters in the group law to reinforce the fact that all
physical variables do emerge naturally from the group manifold itself and
the dimension will be kept to $1+1$ to reduce the expressions to the
minimum.

Reading the group law (\ref{Galileo}) in the new variables and deriving
the double primed variables with respect to every non-primed and primed
one at the identity we get the explicit expressions of the left- and right-vector 
fields, respectively:

\be \begin{array}{ll} \tilde{X}^L_t=\frac{\!\!\partial}{\partial
t}+v\frac{\!\!\partial} {\partial x}
 +\frac{1}{2}mv^2\frac{\!\!\partial}{\partial\phi} &
 \tilde{X}^R_t=\frac{\!\!\partial}{\partial t}\\
\tilde{X}^L_x=\frac{\!\!\partial}{\partial x} & \tilde{X}^R_x=
\frac{\!\!\partial}{\partial x}
 +mv\frac{\!\!\partial}{\partial\phi}\\
\tilde{X}^L_v= \frac{\!\!\partial}{\partial v}
 +mx\frac{\!\!\partial}{\partial\phi} &  \tilde{X}^R_v=
\frac{\!\!\partial}{\partial v}+t\frac{\!\!\partial}{\partial x}
 +mtv\frac{\!\!\partial}{\partial\phi}\\
 \tilde{X}^L_\phi= \frac{\!\!\partial}{\partial\phi}
 &  \tilde{X}^R_\phi= \frac{\!\!\partial}{\partial\phi}
\end{array}\ee

By duality on the left generators, selecting the $U(1)$ component, or by
using the direct formula
\be\Theta\equiv\theta^{L\phi}=\frac{\partial\phi''}{\partial g}|_{g'=g^{-1},g)}\,,\label{thetapo}\ee

\noi one can compute the quantization form (the actual expression of the
Poincar\'e-Cartan part is defined up to a total differential depending
of the particular co-cycle used in the group law, which is defined in
turns up to a co-boundary; see Ref.\cite{Bargmann,23}):
\[  {\Theta\equiv{\theta^L}^\phi=-mxdv-\frac{1}{2}mv^2dt+d\phi} \]
{}From the commutation relations (the left ones change the structure
constants by a global sign)
\be \left[\tilde{X}^R_t,\;\tilde{X}^R_x\right]=0, \;\;
\left[\tilde{X}^R_t,\;\tilde{X}^R_v\right]=\tilde{X}^R_x, \;\;
\left[\tilde{X}^R_x,\;\tilde{X}^R_v\right]={-m\tilde{X}^R_\phi}\ee
\noi one rapidly identifies $x,v$ as canonically conjugated (symplectic)
variables and $t$ as a non-symplectic parameter. In fact, the left generator
$\tilde{X}^L_t$ generates the characteristic subalgebra ${\cal G}_\Theta$
and constitutes the classical equations of motion (generalized, 
since there is an extra equation for the central parameter).

The quantum wave functions are complex functions on $\tilde{G}$,
$\Psi=\Psi(\zeta,x,v,t)$, restricted by the $U(1)$-function condition
(\ref{u1function}), as well as the polarization conditions
\[\tilde{X}^L_a\Psi=0 \;\;\;(a=t,x \; \hbox{maximal set}) \]

\noi We then obtain:
\bea \tilde{X}^L_\phi\Psi&=&i\Psi\Rightarrow\;\;\Psi=\zeta\Phi(t,x,v)\nn\\  
\tilde{X}^L_x\Psi&=&0\Rightarrow\;\;\Phi\neq\Phi(x),\;\Phi=\varphi(t,v)\nn\\ 
 \tilde{X}^L_t\Psi&=&0\Rightarrow\;\;\frac{\partial\varphi}{\partial 
t}+\frac{i}{2} mv^2\varphi=0\;\Rightarrow\; 
{i\frac{\partial\varphi}{\partial t}=\frac{p^2}{2m}\varphi}\; ,\nn \eea 

\noindent i.e. the Schr\"odinger equation in momentum space. 

On the (reduced) wave functions the right-invariant vector fields act
reproducing the standard quantum operators in momentum space
``representation":
\be \tilde{X}^R_x\varphi=imv\varphi,\;\;
\tilde{X}^R_v\varphi=\frac{\!\partial}{\partial v}\varphi,\;\;
\tilde{X}^R_t\varphi=-i\frac{p^2}{2m}\varphi,\ee

\noi the operator $\hat{E}\equiv i\tilde{X}^R_t$ being a function of the
basic one $\hat{p}\equiv i\tilde{X}^R_x$. Had we considered the motion in
$3+1$ dimensions, we would have found also new operators in the
characteristic subalgebra associated with rotations acquiring the usual
expressions in terms of the basic operators $\hat{\vec{v}}$ and
$\hat{\vec{x}}\equiv i\tilde{X}^R_{\vec{v}}$.

\subsection{Revisited Minimal Coupling Principle}
In this subsection we attempt to describe group-theoretically the motion
of a particle subject to an external field. Even though we do not intend
to account for the field degrees of freedom, the transformation properties
of its ``zero-modes" can be encoded into part of a symmetry group. The
general mechanism under which a free particle starts suffering an
interaction parallels the well-known Minimal Coupling Principle, which is
now revisited from our group-theoretical approach. We shall be concerned
here with the classical domain only.

Let $\tilde{G}$ be a quantization group generated by $\{\tilde X_A\},
\;A=1,...,n$ and $\{\tilde X_a\},\;a=1,...,m<n$ an  {invariant} subalgebra:

\[ [\tilde X_A,\;\tilde X_{ {a}}]=C^{ {b}}_{A {a}}
\tilde X_{ {b}}\]

\noi If we make  {``local"} the subgroup generated by $\{\tilde X_{ {a}}\}$, that
is to say, if the corresponding group variables are allowed to depend
arbitrarily on the space-time parameters, we get an infinite-dimensional
Lie algebra:

\[ \{f^{ {a}}\otimes \tilde X_{ {a}},\;\tilde X_A\}\]

\noi with the following new commutators:
\bea \left[\tilde X_A,\;f^{ {a}}\otimes \tilde X_{ {a}}\right]&=& f^{
{a}}\otimes\left[\tilde X_A,\;\tilde X_{ {a}}\right]+ L_{\tilde X_A}f^{ {a}}\otimes \tilde X_{ 
{a}}\nn \\ &=& f^{ {a}}\otimes C^{ {b}}_{A {a}} \tilde X_{ {b}}+L_{\tilde X_A}f^{ {a}}\otimes \tilde X_{ 
{a}}\label{localalgebra} \eea

\noi Now we just  attempt to ``quantize" this new (local) group $
{\tilde{G}(\vec{x},t)}$.

\subsubsection{Particle in an Electromagnetic Field.}
We start from the $U(1)$-extended Galilei group, $\tilde{G}$, and make the
rigid group $\zeta=e^{i\phi}\in U(1)$ into  ``local", i.e. we allow the
parameter to depend on the space-time variables, $\phi=\phi(\vec{x},t)$. The idea is
to keep the invariance of the generalized Poincar\'e-Cartan form
$\Theta=p_idx^i-\frac{\vec{p}\,^2}{2m}dt+d\phi$ under the locally extended
Galilei group.

According to the  {\it Revisited Minimal Coupling Principle} \cite{electrograv} we only
have to compute the 1-form $\Theta$ associated with the Galilei group
extended by the infinite dimensional group $U(1)(\vec{x},t)$. But, in
order to parameterize properly the quantization group let us formally
write
\be
\phi(\vec{x},t)=\phi(0,0)+\phi_\mu(\vec{x},t)x^\mu \equiv\phi+ 
{A_\mu(\vec{x},t)}x^\mu\nn \ee

\noi and compute the group law:
\\

$
\begin{array}{lcl}
t''&=&t'+t\nn\\
\vec{x}''&=&\vec{x}'+R'\vec{x}+\vec{v}\,'t\nn\\
\vec{v}''&=&\vec{v}\,'+R'\vec{v}\nn\\
A_{\vec{x}}''&=&A_{\vec{x}}'+R'A_{\vec{x}}\nn\\
A_t''&=&A_t'+A_t+\vec{v}\,'\cdot R'A_{\vec{x}}\nn\\
\phi''&=&\phi'+\phi+ {\bf m}[\vec{x}\,'\cdot R'\vec{v}+t(\vec{v}\,'\cdot
R'\vec{v}+\frac{1}{2}v'^2)]+{\bf q}[\vec{x}\,'\cdot R'A_{\vec{x}}+t\vec{v}\,'\cdot R'A_{\vec{x}}+tA_t']\nn
\end{array}
$\\

\noi where two different co-cycles characterized by $m$ and $q$, that is,
the mass and the electric charge, have been introduced.

From now on we shall disregard the rotation subgroup although the vector
character of the variables will be maintained. Also, and since we do not
intend to describe quantum aspects, the expression of the left-invariant
generators will be omitted (see Ref. \cite{electrograv}) and only  the Lie
algebra commutators are
written:\\
\begin{equation}
\begin{array}{lll}
\left[\tilde{X}^L_t, \tilde{X}^L_{\vec{x}} \right]=0 &
\left[\tilde{X}^L_{x^i}, \tilde{X}^L_{A_{x^j}} \right]=
{\bf q}\delta_{ij}\tilde{X}^L_\phi & \left[\tilde{X}^L_t,
\tilde{X}^L_{\vec{v}} \right]= -\tilde{X}^L_{\vec{x}}
\\ \left[\tilde{X}^L_{\vec{x}}, \tilde{X}^L_{A_t} \right]= 0 &
\left[\tilde{X}^L_t, \tilde{X}^L_{A_x} \right]= 0 &
\left[\tilde{X}^L_v, \tilde{X}^L_{A_x} \right]=\tilde{X}^L_{A_t}\\
\left[\tilde{X}^L_t, \tilde{X}^L_{A_t} \right]=- {\bf q}\tilde{X}^L_\phi &
\left[\tilde{X}^L_v, \tilde{X}^L_{A_t} \right]=0 &
\left[\tilde{X}^L_{x^i}, \tilde{X}^L_{v^j} \right]=
{\bf m}\delta_{ij}\tilde{X}^L_\phi
\end{array}\label{electroalgebra}
\end{equation}

 \noi By duality from the explicit expression of the left-invariant generators,
derived in turn from the group law, or directly from the composition law
corresponding to the $U(1)$ parameter, through the formula
(\ref{thetapo}), we obtain the quantization form
\bea \Theta=- {m}\vec{x}\cdot d\vec{v}- {q}\; \vec{x} \cdot
d\vec{A}-(\frac{1}{2} {m}\vec{v}\,^2+ {q} A_t)dt+d\phi \nn \eea

\noi whose characteristic module contains the generator of the time evolution: $X$
such that \ \ \  $i_X\Theta=i_Xd\Theta=0$, that is,
\[
X=\frac{\!\!\partial}{\partial t}+\vec{v}\cdot\frac{\!\!\partial}{\partial
\vec{x}} -\frac{ {q}} { {m}}\left[ (\frac{\partial A_i} {\partial
x^j}-\frac{\partial A_j}{\partial x^i})v^j+\frac{\partial A_0} {\partial
x^i}+\frac{\partial A_i}{\partial t}\right] \frac{\!\!\partial}{\partial v_i}
\]

\noi which implies the following explicit equations of motion:
\be \frac{d\vec{x}}{dt}=\vec{v},\;\; {m}\frac{d\vec{v}}{dt}={q}
[\vec{v}\wedge(\vec{\nabla}\wedge\vec{A})-\vec{\nabla}A_0-\frac{\partial
\vec{A}}{\partial t}]\,. \label{Lorentzforce} \ee

Making the standard change of variables
\be \vec{B}\equiv\vec{\nabla}\wedge\vec{A},\;\;\;
\vec{E}\equiv-\vec{\nabla}A_0-\frac{\partial \vec{A}}{\partial t} \ee

\noi we finally arrive at the ordinary equation of a particle suffering
the Lorentz force:
\[{m}\frac{d\vec{v}}{dt}= {q}[\vec{E}+\vec{v}\wedge\vec{B}]\]

As a last general comment, let us remark once again the physical relevance
of central extensions. It might seem paradoxical the fact that a
non-trivial vector potential (in the sense that it is not the gradient of
a function) can be derived some how from the function $\phi(\vec{x},t)$,
but it is the central extension mechanism what insures that $A_\mu$ can be
something different from the gradient of a scalar function. In other
words, the generator $X_{A_\mu}$ in (\ref{electroalgebra}), for $q=0$,
necessarily generates trivial (gauge) changes in $A_\mu$.

\subsubsection{Particle in a gravitational field.}
Let us pass very briefly through this finite-dimensional example
where the computations are made in dimension $1+1$ although the vector
notation is restored at the end. We start from the $U(1)$-extended
Poincar\'e group and make ``local"
the translation subgroup, the Lie algebra of which can be written as\\

$\begin{array}{lcl}
\left[\tilde{X}^R_t,\;\tilde{X}^R_x\right]=0 &&\;\;\;\;\;\left[P_0,P\right]=0\\
\left[\tilde{X}^R_t,\;\tilde{X}^R_v\right]=\tilde{X}^R_x
&\;\;\;\hbox{or}&\;\;\;\;\;
\left[P_0,K\right]=P\\
\left[\tilde{X}^R_x,\;\tilde{X}^R_v\right]=-\frac{1}{c^2}\tilde{X}^R_t-
 {m}\tilde{X}^R_\phi&&
\;\;\;\;\;\left[P_0,K\right]=-\frac{1}{c^2}P_0- {m}X_\phi\,,
\end{array}$\\

\noi and repeat the process of ``localizing" translations in a way
analogous to that followed for the $U(1)$ subgroup in the electromagnetic
case. We write, as before,
\[  {f^\mu}\otimes
P_\mu=(f^\mu(0)+ {f^{\mu\sigma}(x)}x_\sigma ) \otimes P_\mu \]

\noi and rename the functions $f^{\mu\nu}$ as $h^{\mu\sigma}$, which will
prove to be the non-Minkowskian part of a non-trivial metric, that is:
$h^{\mu\nu}\equiv g^{\mu\nu}\;-\;\eta^{\mu\nu}$

The Lie algebra must be explicitly written according to the general
formula (\ref{localalgebra}) and the rigid algebra (extended
Poincar\'e). We show in boldface the terms that survive after an
In\"on\"u-Wigner contraction with respect to the subgroup generated by
$\t$ (the non-relativistic
limit):
\\

$
\begin{array}{lll}
\left[\v, \x\right] = -\t + {\bf mc\z} & \left[\t,\htx\right] = {\bf \x} & 
\left[\v, \t\right] = {\bf -\x} \\ \left[\x,\htx\right] = -\t {\bf - g\z} 
& \left[X_v, \htt\right]= -\htx & \left[\x,\hxx\right] = -\x\\ 
\left[\htt,\htx\right] = \v & \left[X_v, \hxx\right]= \htx & 
\left[\htx,\hxx\right] = \v \\ \left[\t,\htt\right] = \t + {\bf g\z} & 
\left[X_v, \htx\right]= -\htt + \hxx 
\end{array}$
\\

It should be remarked that we have naively written a {\it gravitational
coupling constant} {\bf $g$} in places that parallel those of the electric
charge {\bf $q$} in the Lie algebra that accounts for the electromagnetic
interaction; that is to say, on the right hand side of the commutators
$\left[\t,\htt\right]$ and $\left[\x,\htx\right]$, but the Jacobi identity
requires the equality {\bf $g=mc$}, which may be properly identified with an
algebraic version of  the {\bf Equivalence Principle}. Note that both $q$
and $g=mc$ are true central charges (in the sense that they parameterize non-trivial central extensions) in the
non-relativistic limit. To be precise, $q$ also parameterizes non-trivial central extension in the Poincar\'e group.

Now, the group law must be computed order by order, although it is enough
to keep the expansion up to the  $3^{th}$ order for illustrating the dynamics. We
remit the readers to Ref.~\cite{electrograv} for a detailed computation
and here only the final equation of motion are showed.

\noindent {\bf Geodesic Force:} We introduce for simplicity the vector
notation: $ h^{0i}\equiv\vec{h}$. In terms of these variables the
equations of motion, for low gravity and low velocity, are:
\be \frac{d\vec{x}}{dt}=\vec{v},\;\;\;  m\frac{d\vec{v}}{dt}=-
{m\left[\vec{v}\wedge(\vec{\nabla}\wedge
    \vec{h})-\vec{\nabla}h^{00}-\frac{\partial\vec{h}}{\partial t}\right]+\frac{m}{4}
    \vec{\nabla}(\vec{h}\cdot\vec{h})}\ee

They reproduce the standard geodesic motion, up to the limits mentioned,
and in a form that emulate the electromagnetic motion (\ref{Lorentzforce})
for electromagnetic-like vector potential ${\cal
A}=(h^{00}-\frac{1}{4}\vec{h}\cdot\vec{h},\;\vec{h})$ according to that
which is named ``gravitoelectromagnetic" description in the literature
(see, for instance, Ref. \cite{Wald,Tucker,Maartens,Mashhoon}).

\subsection{Particle moving on a group manifold: case of the $SU(2)$ group}

In this last finite-dimensional example let us adopt a slightly different point of view, that is, 
we shall start from the classical Lagrangian and try to close a
Poisson subalgebra containing $ \langle H,q,p\rangle$. However, except for simple examples such 
a subalgebra is infinite. To keep ourselves in finite dimensions (that is, with a finite number
of degrees of freedom) we may alternatively resort to 
an auxiliary, different finite-dimensional Poisson subalgebra (closing a group $G$) such that, in its enveloping 
algebra, the original functions $ \langle H,q,p\rangle$ can be found, and therefore quantized. 
In fact, in the GAQ scheme, not only the generators of the original group $G$
can be quantized, but also the entire universal enveloping algebra. This procedure 
has been explicitly achieved in dealing with the quantum dynamics of a particle in a 
(modified) P\"oschl-Teller potential\cite{Pasteles}, where the ``first-order" (auxiliary) group 
$G$ used was $SL(2,R)$.

We start by parametrizing  rotations with a vector $\vec{\varepsilon}$ in the rotation-axis direction and with 
modulus
\[|\vec{\varepsilon}|=2\hbox{sin}\frac{\varphi}{2}\]
\[R(\vec{\varepsilon})^i_j=(1-\frac{\vec{\varepsilon}\,^2}{2})\delta^i_j-
\sqrt{1-\frac{\vec{\varepsilon}\,^2}{4}}\eta_{\cdot jk}^{i}\varepsilon^k+\frac{1}{2}\varepsilon^i\varepsilon_j\]

In these coordinates the canonical left-invariant $1$-forms read:
\[ \theta^{L(i)}_j=\left[\sqrt{1-\frac{\vec{\varepsilon}\,^2}{4}}\delta^i_j+\frac{\varepsilon^i\varepsilon_j}
{4\sqrt{1-\frac{\vec{\varepsilon}\,^2}{4}}}+\frac{1}{2}\eta_{.jm}^i\varepsilon^m\right] \]

\noi and in terms of these the {\it particle-$\sigma$-Model Lagrangian} acquires the following expression:
\[{\cal L}= 
\frac{1}{2}\delta_{ij}\theta^{L(i)}_m\theta^{L(j)}_n\dot{\varepsilon}^m\dot{\varepsilon}^n=
\frac{1}{2}\left[\delta_{ij}+\frac{\varepsilon_i\varepsilon_j}{4(1-\frac{\vec{\varepsilon}\,^2}{4})}\right]
\dot{\varepsilon}^i\dot{\varepsilon}^j\equiv\frac{1}{2}g_{ij}\dot{\varepsilon}^i\dot{\varepsilon}^j \]

 Proceeding much in the same way followed in the previous section, we compute the canonical momenta: 
 \[\pi_i=\frac{\partial{\cal L}}{\partial\dot{\varepsilon}^i}=g_{ij}\dot{\varepsilon}^j\]

\noi and the Hamiltonian: 
\[{\cal H}=\pi_i\dot{\varepsilon}^i-{\cal L}=\frac{1}{2}g^{-1ij}\pi_i\pi_j\]

We assume the canonical bracket between the basic functions $\varepsilon^i$ and $\pi_j$: 
\bea
\{\varepsilon^i,\,\pi_j\}&=&\delta^i_j \ \ \ \hbox{added with}\nn\\
\{{\cal H},\,\varepsilon^i\}&=&-g^{-1ij}\pi_j\nn\\
\{{\cal H},\,\pi_i\}&=&\frac{1}{2}(\vec{\varepsilon}\cdot\vec{\pi})\pi_i\;,\nn
\eea

\noi so that $ \langle {\cal H},\,\varepsilon^i,\,\pi_j,\,1\rangle$ 
{\it do not close} a finite-dimensional Lie algebra.

However, we may define the following set of new {\it ``coordinates"}, {\it ``momenta"} and even, 
{\it ``energy"} and {\it ``angular momenta"}:
\[ \langle \,p^i\equiv 
2g^{-1ij}\pi_j,\;k^j\equiv\sqrt{2{\cal H}}\varepsilon^j,\;E\equiv 
2\sqrt{2{\cal H}},\;J^k\equiv\eta_{\cdot mn}^k\varepsilon^m\pi^n\,\rangle\;.\]

\noi They close the Lie algebra of $SO(3,2)$ i.e. an Anti-de Sitter 
algebra. That is, the basic brackets:
\bea
\{E,\,p^i\}&=&k^i\nn\\
\{E,\,k^j\}&=&-p^j\nn\\
\{k^i,\,p_j\}&=&\delta^i_jE\,,\nn
\eea

\noi along with the induced ones:
\bea
\{k^i,\,k^j\}&=&{\bf -}\eta^{ij\cdot}_{\;\;k}J^k\;\;\;\{p^i,\,p^j\}\,=\,{\bf -}\eta^{ij\cdot}_{\;\;k}J^k\nn\\
\{J^i,\,J^j\}&=&\eta^{ij\cdot}_{\;\;k}J^k\;\;\;\;\;\;\{J^i,\,k^j\}\,=\,\eta^{ij\cdot}_{\;\;k}k^k\nn\\
\{J^i,\,p^j\}&=&\eta^{ij\cdot}_{\;\;k}p^k\;\;\;\;\;\;\{E,\,\vec{J}\}\,=\,0,\nn
\eea

\noi close a finite-dimensional Lie algebra to which we may apply the GAQ.
 ({\it Note} the minus sign in the first line, which states that the involved group is 
$SO(3,2)$ and not $SO(4,1)$).

We then quantize the Anti-de Sitter group so that the original operators 
$\hat{{\cal H}},\;\hat{\pi}_i,\;\hat{\varepsilon}^j$ can be found in its {\it enveloping algebra}
through the expression: 
\[ 
\langle \hat{\cal H}\equiv \frac{1}{8}\hat{E}^2, \, 
\hat{\pi}^i\equiv \frac{1}{4} \left(\hat{g}^{ij} \hat{p}_j + \hat{p}_j 
\hat{g}^{ij}\right),\, 
\hat{\varepsilon}^i\equiv \frac{1}{2\sqrt{2}} 
\left(\hat{\cal H}^{-1/2} \hat{k}^i + \hat{k}^i \hat{\cal H}^{-1/2}\right),\, 
\hat{1}\rangle
\]

In so doing we parameterize a central extension of the Anti-de Sitter group by (abstract) 
variables $\{a^0,\,\vec a,\,\vec \nu,\,\vec \epsilon,\,\zeta\}$, which 
mimic those for the Poincar\'e group. In the same way we hope that the 
right-invariant generators associated with those parameters reproduce corresponding functions 
$\{E,\, p_i,\, k^j,\, J^k,\, 1\}$ as Noether invariants satisfying the Poisson brackets above.  

At this point it should be stressed that the parameters $\vec \epsilon$ and 
the corresponding quantum operators (essentially the right generators $\tilde{X}^R_{\vec 
\epsilon}$) are associated with ordinary rotations on Anti-de Sitter space-time, 
whereas the $\vec \varepsilon$ parameters correspond to ``translations'' on the $SU(2)$ manifold.

We shall not give here the explicit group law for the group variables nor the explicit expression 
for the left-invariant vector fields on the extended $SO(3,2)$ group
(which can be found in Refs. \cite{silvester,needs}), limiting ourselves to write
the explicit expression for the (higher-order) polarization condition and the
wave functions as solutions of it. That is, from the polarization condition 
$\mathcal P \psi = 0$,  with
\[{\cal 
P}= \langle \,\tilde{X}^{LHO}_{a^0}\equiv(\tilde{X}^L_{a^0})^2-c^2(\tilde{X}^L_{\vec{a}})^2-
\frac{2imc^2}{\hbar}\tilde{X}^L_{a^0},\;\tilde{X}^L_{\vec{\nu}},\;\tilde{X}^L_{\vec{\epsilon}}\,\rangle\]

\noi we arrive at a 
wave function depending only on $a^0,\, \vec a$ and satisfying a Klein-Gordon-like 
equation with $SO(3,2)$ D'Alembertian operator given by
\bea
\square & = & 
\frac{1}{16 q_{a}^2}\{
[16-(a^0)^2\frac{\omega^2}{c^2} (8+\frac{\omega^2}{c^2}(r_a^2-(a^0)^2))]  \frac{\!\partial^2}{\partial a^{02}}
- a^0\frac{\omega^2}{c^2}[40+7\frac{\omega^2}{c^2}(r_a^2 \nn \\
&-&(a^0)^2)]\frac{\!\partial}{\partial a^0}
%
 -[16 +r_a^2 \frac{\omega^2}{c^2}(8+\frac{\omega^2}{c^2}(r_a^2-(a^0)^2))]\frac{\!\partial^2}{\partial r_a^2}
 - \frac{1}{r_a}[32+\frac{\omega^2}{c^2}(40\nn\\
 &+&7\frac{\omega^2}{c^2}(r_a^2-(a^0)^2))]\frac{\!\partial}{\partial r_a} 
2 a^0 r_a \frac{\omega^2}{c^2}(8+\frac{\omega^2}{c^2}(r_a^2-(a^0)^2))
\frac{\!\partial^2}{\partial a^{0}\partial r_a}\}
+ \frac{\vec{L}^2}{q_a^2 r_a^2}\, ,   \nn
\eea

\noi where 
\[\vec{L}^2= 
-\frac{1}{\hbox{sin}\theta_a}\frac{\!\partial}{\partial\theta_a}(\hbox{sin}\theta_a\frac{\!\partial}{\partial\theta_a})-
\frac{1}{\hbox{sin}^2\theta_a}\frac{\!\partial^2}{\partial\varphi_a^2}
\]

\noi is the square of the standard orbital angular momentum operator (save for a factor $\hbar$)
 and 
\be
r_a = \sqrt{\vec a \cdot \vec a}\;,\; q_{a}\,=\,\sqrt{1+\frac{\omega^2}{4 c^2}\, (\vec a^{\,2} 
-(a^0)^2)}\,. \nn
\ee

The {\it wave functions} are 
\[ \phi(\vec{a},a^0)= 
e^{-2ic\lambda_{nl} 
\arcsin(\frac{\omega q_{a} a^0}{\sqrt{4c^2+\omega^2 q_{a}^2 r_a^{2}}})}\;
Y^l_m(\theta_a,\varphi_a)(1+\frac{\omega^2}{c^2}q_a^2 r_a^{2})^{-\frac{\lambda_{nl}}{2}}(q_a r_a)^l\phi^{\lambda_{nl}}_l(q_a r_a)
\]
\noi where 
\bea
\lambda_{nl}&\equiv&\frac{E}{\hbar\omega}\nn\\
E&\equiv&(\frac{3}{2}+2n+l+\frac{1}{2}\sqrt{9+4\frac{m^2c^2}{\hbar^2\omega^2}-48\xi})\hbar\omega\nn\\
\phi^{\lambda_{nl}}_l&=&{}_2F_1(-n,n+l+\frac{3}{2}-\lambda_{nl},l+\frac{3}{2};-\frac{\omega^2}{c^2}q_a^2 r_a^2)\nn
\eea

\noi and $\xi$ is a free parameter related to the {\it ``zero-point 
energy"} \cite{Pepe2}. On this representation the operators corresponding to the original functions 
$\varepsilon^i,\;\pi_j$ and the energy $\cal H$ can be realized.

\section{Quantum Field Theory (examples with an infinite number of degrees of freedom)}
Typical infinite-dimensional systems in Physics appear as mappings from a
space-time manifold $M$ into a (not necessarily Abelian) target group $G$
\be  {\varphi: M \rightarrow  G}, \;\; x\mapsto \varphi(x).\label{mapp}\ee

If $a$ is an invertible, differentiable transformation of $M$, i.e. $a$ is
an element in Diff$(M)$, or a subgroup of it, the following semi-direct
$({{\rm Diff}(M)\otimes_s G(M)})$ group  law holds:
\be a''=a'\circ a,\;\;\;\;
\varphi''(x)=\varphi'(a(x))*\varphi(x),\label{gaugegroup} \ee
\noi where $\circ$ is the composition group law in Diff$(M)$ (composition
of mappings), $*$ denotes the composition group law in the target group
$G$ and $a(x)$ stands for the action of Diff$(M)$ on $M$. When the group
$G$ is not a (complex) vector space $\mathbb C^n$, the group of mappings
is usually called {gauge}, local or current group $G(M)$. Specially
well-known are the unitary gauge groups on Minkowski space-time and the
loop groups which correspond to the case in which $M$ is the circle $S^1$.
However, the actual physical fields correspond to the elements in the
centrally extended group $\widetilde{{\rm Diff}(M)\otimes_s G(M)}$. In the
case of $M=S^1$ the group (\ref{gaugegroup}) has a specially rich
structure (that is Virasoro$\;\otimes_s\;$Kac-Moody) \cite{Milikito} with
many applications in  conformal field theory in $1+1$ dimensions
\cite{Olive}. As a general comment, the ability in parametrizing the
infinite-dimensional group (\ref{gaugegroup}) will play a preponderant
role in the corresponding physical description.

\subsection{The Klein-Gordon Field}

As a very simple example of the general scheme above-mentioned let us
consider the case in which $M$ is the Minkowski space-time with co-ordinates
 $(x^0\equiv ct,\, \vec{x})$, Diff$(M)$ is restricted to its
Poincar\'e subgroup (or even just the space-time translations subgroup,
for the sake of simplicity), parameterized by $(a^0\equiv cb,\, \vec{a})$,
and $G$ is simply a complex vector space, let us say  $\mathbb C$,
parameterized by $\varphi$.

There is a natural parametrization of the group above associated with
a factorization of $M$ as $\Sigma\times R$, that is, the product of a Cauchy surface
$\Sigma$ and the time real line. In fact, we can use $\langle
b,\vec{a};\varphi(\vec{x}),\dot{\varphi}(\vec{x})\rangle$. In these
variables, however, whereas the space translations $\vec{a}$ act on
$\varphi(\vec{x})$ by just moving the arguments as $\vec{x}+\vec{a}$:
$\varphi(\vec{x}+\vec{a})=\exp(i\vec{a}\cdot\partial_{\vec{x}})\varphi(\vec{x})$,
making $b$ an action on $(\varphi(\vec{x}),\dot{\varphi}(\vec{x}))$
{requires} the knowledge of the equation of motion, though not necessarily
their solutions. For the Klein-Gordon field the time evolution equations
are
\[ {\stackrel{..}{\varphi}(\vec{x})=(\vec{\nabla}^2-m^2)\varphi(\vec{x})}\]

\noi and the time action  {$\varphi'(b(\vec{x}))$} reads:

\[
 {\varphi'( {b}(\vec{x}))\equiv e^{i {b}c\partial_0}\varphi'(\vec{x})=\cos\left[ {b}c\sqrt{m^2-\vec{\nabla}^2}\right]\varphi'(\vec{x})+
i\frac{\hbox{sin}\left[
{b}c\sqrt{m^2-\vec{\nabla}^2}\right]}{\sqrt{m^2-\vec{\nabla}^2}}\dot{\varphi}'(\vec{x})}
\]

\noi This way, all canonical operations on groups can be easily performed.
For instance, the  {right-invariant vector fields} are:
\bea
X^R_b&=&\frac{\!\partial}{\partial b}\nn\\
X^R_{\varphi(\vec{x})}&=&\hbox{cos}\left(b\sqrt{m^2-\vec{\nabla}^2}\right)\frac{\!\delta}{\delta\varphi(\vec{x})}
    -\sqrt{m^2-\vec{\nabla}^2}\hbox{sin}\left(b\sqrt{m^2-\vec{\nabla}^2}\right)\frac{\!\delta}{\delta\dot{\varphi}(\vec{x})}\nn\\
X^R_{\dot{\varphi}(\vec{x})}&=&\hbox{cos}\left(b\sqrt{m^2-\vec{\nabla}^2}\right)\frac{\!\delta}{\delta\dot{\varphi}(\vec{x})}
    +\frac{1}{\sqrt{m^2-\vec{\nabla}^2}}\hbox{sin}\left(b\sqrt{m^2-\vec{\nabla}^2}\right)\frac{\!\delta}{\delta\varphi(\vec{x})},\nn
\eea
and their commutation relations: \bea
\left[X^R_b,\,X^R_{\varphi(\vec{x})}\right]&=& {-(m^2-\vec{\nabla}^2)}X^R_{\dot{\varphi}(\vec{x})}\nn\\
\left[X^R_b,\,X^R_{\dot{\varphi}(\vec{x})}\right]&=&X^R_{\varphi(\vec{x})}\nn\\
\left[X^R_{\varphi(\vec{x})},\,X^R_{\dot{\varphi}(\vec{x}')}\right]&=&0
\;\;\;(\; \delta(\vec{x}-\vec{x}')X^R_\phi\;\; \hbox{when centrally
extended}).\nn \eea

\noi Notice that the actual solutions of the equations of motion of a more
general system are not required since the corresponding Lie algebra  can
be exponentiated (at least) order by order giving rise to the finite
action of  $b$ on both $\varphi(\vec{x})$ and $\dot{\varphi}(\vec{x})$.

As mentioned above, what really matters for the physical description is
the corresponding centrally extended group. In order to motivate such
extension we shall proceed in a way analogous to that followed in the case
of Mechanics. Let us  go then temporarily to the standard Lagrangian
formalism for classical fields. The real Klein-Gordon field of mass $m$ is
described by the Lagrangian
\be {\cal L} =
\frac{1}{2}(\partial_\mu\varphi\partial^\mu\varphi-m^2\varphi^2)\label{Klein-Gordon}
\ee

\noi which is well-known to realize the Poincar\'e symmetries (see, for
instance Ref. \cite{Itzykson}). However, the Noether invariants associated
with space-time symmetries are not relevant in studying the solution
manifold ${\cal M}$ in the sense that they are not the basic, independent
functions parametrizing the phase space. In fact, quantities such as the
energy-momentum tensor or the generalized (rotations and Lorentz) angular
momenta are written in terms of the Fourier coefficients $a(k),\;a^*(k)$,
where the four vector $k_\mu$ runs on the Lorentz orbit $k^\mu k_\mu =
m^2$. Here we are primarily interested in characterizing those Fourier
coefficients as Noether invariants of certain generators leaving
semi-invariant the Lagrangian (\ref{Klein-Gordon}). To this end we consider
the following vector fields on the complete (including the field
derivatives) configuration space for the Klein-Gordon Field ($x^\nu,
\varphi, \varphi_\mu$):
\be \bar{X}_{a^*(k)}\equiv ie^{ikx}\frac{\!\partial}{\partial\varphi}
-k_\nu ie^{ikx}\frac{\!\partial}{\partial\varphi_\nu} \ee

Computing the Lie derivative of the Lagrangian with respect to this vector
(note that the second components of this vector are simply the derivatives
of the components on $\varphi$) we obtain:
\be L_{\bar{X}_{a^*(k)}}{\cal L}= \partial_\mu\beta^\mu, \;\;
\beta^\mu\equiv -k^\mu e^{ikx}\varphi, \label{semiinv}\ee

\noi where explicit use  of the mass-shell condition for $k$ has been
made. The Noether theorem establishes that the current
\[ J^\mu_{a^*(k)}= X^\varphi_{a^*(k)}\pi^\mu-\beta^\mu \]

\noi (where $X^\varphi_{a^*(k)}$ is the $\varphi$-component of the
generator, i.e. the infinitesimal variation $\delta\varphi$ and
$\pi^\mu\equiv\frac{\partial{\cal L}}{\partial\varphi_\mu}$ is the field
covariant momentum) is conserved: $\partial_\mu J^\mu=0$. The Noether
charge reads:
\be Q_{X_{a^*(k)}}=\int d^3xJ^0=i\int
d^3xe^{ikx}(\dot{\varphi}-ik^0\varphi)\label{Noechar}\ee

\noi which turns out to be just the Fourier coefficient $a(k)$. In the
same way we obtain the charge $a^*(k)$ and so a coordinate system for the
solution manifold ${\cal M}$ made of Noether invariants.

Analogously, an equivalent configuration-space parametrization can be
considered. In fact, the vector fields:
\bea X_{\pi(\vec{y})}&\equiv&
i\int\frac{d^3k}{2k^0}\left[e^{i\vec{k}\cdot\vec{y}}e^{ikx}\frac{\!\partial}{\partial\varphi}-h.c\right]\label{semi-symmetry}\\
X_{\varphi(\vec{y})}&\equiv&
-\int\frac{d^3k}{2k^0}k^0\left[e^{i\vec{k}\cdot\vec{y}}e^{ikx}\frac{\!\partial}{\partial\varphi}+
h.c\right]\nn \eea

\noi have as Noether invariants the values of $\varphi$ and
$\pi=\dot\varphi$ on each one of the points of the Cauchy surface
$\Sigma$. In terms of these ``configuration-space" variables the
symplectic form on ${\cal M}$ adopts the aspect and properties of that of
Classical Mechanics:
\[ w_{K-G}=d\vartheta_{K-G}=d\left(\int d^3x\;\pi(\vec{x})\delta\varphi(\vec{x})\right)\,,\]

\noi the symplectic potential $\vartheta_{K-G}$, or Liouville form, being {\it
semi-invariant under the basic symmetries} (\ref{semi-symmetry}).

To end up with the (semi-)invariance properties of the Klein-Gordon field
it should be mentioned that the symmetries of this Lagrangian can be given
the aspect of some sort of ``residual gauge" symmetry, even for $m\neq 0$.
In fact, for any real function $f$ on $M$ satisfying the Klein-Gordon
equation, the vector field
\[
X^f=f\frac{\!\partial}{\partial\varphi}+f_\mu\frac{\!\partial}{\partial\varphi_\mu}\]

\noi leaves (\ref{Klein-Gordon}) semi-invariant.

Once the necessity of a central extension of the semi-invariance group
(\ref{gaugegroup}) has been stated and motivated,  we write the
quantization group for the Klein-Gordon field in covariant form
\cite{Miguel} as follows:
\bea
a''&=&a'+\Lambda'a\nn\\
\Lambda''&=&\Lambda'\Lambda\nn\\
\varphi''(x)&=&\varphi'(\Lambda x+a)+\varphi(x)\nn\\
\varphi_\mu''(x) &=&\varphi_\mu'(\Lambda x+a)+\varphi_\mu(x)\nn\\
\zeta''&=&\zeta'\zeta \exp\left\{\frac{i}{2}\int_\Sigma
d\sigma^\mu\left[\varphi'(\Lambda x+a)\varphi_\mu(x)-\varphi'_\mu(\Lambda
x+a)\varphi(x)\right]\right\}\,,\nn \eea

\noi where the Poincar\'e subgroup is parameterized by $(a,\Lambda)$, 
$d\sigma^\mu = n^\mu(x) d\sigma$ and $x$ is, in principle, supposed to live on the Cauchy 
hypersurface $\Sigma$, with unit normal vector $n^\mu(x)$, although the fields $\varphi$ and $\varphi_\mu$ can be
defined on the entire Minkowski space-time if they satisfy the equations of motion. In this sense,
the exponential is well-defined, even though the fields would live on the whole space-time, because 
the integrand is a conserved current. 
Further details can be found in \cite{Miguel} and references therein.

\subsection{Sigma Model-type systems}

A less trivial example of infinite dimensions arises when the target group,
in which the fields are valued, is considered to be non-Abelian. Then a
physical system associated with such fields is the so-called Non-Linear
Sigma Model (NLSM).

For the sake of simplicity, let us consider the case of $G=SU(2)$.
Denoting by $\tau_a,\, a=1,2,3$, the Lie algebra basis matrices, with commutation relations
$[\tau_a,\tau_b]=\eta_{ab}^{\phantom{ab}c}\tau_c$, and
$U=\exp(i\varphi^b\tau_b)$ an element of $SU(2)$ parameterized by
$\varphi \equiv (\varphi^1,\varphi^2,\varphi^3)$, the left-invariant
canonical 1-form [for the mappings $\varphi^b(x)\in G$ in (\ref{mapp})] is 
$\theta^L_\mu=U^{-1}\partial_\mu U\equiv\theta^{L(b)}_\mu\tau_b$. In terms 
of 
$\theta^{\,L}_\mu\equiv(\theta^{L(1)}_\mu,\theta^{L(2)}_\mu,\theta^{L(3)}_\mu)$,
the Lagrangian of the $SU(2)$ NLSM can be written as:

\begin{equation}
\mathcal L_{\sigma} = \um {\rm Tr}_{G}(\theta^{\,L}_{\mu} \theta^{\,L \mu})
\label{trazatoa}
\end{equation}
This Lagrangian is invariant under the left action (\ref{leftrightact}) of
$SU(2)$ and, eventually, it is right-invariant as well (i.e. it is
chiral).

In facing the group-quantization of these systems we find a serious obstruction:
%
%
It is not possible (without drastically distorting the leading 
$SU(2)$ symmetry) to find a
semi-invariance --like in (\ref{semiinv})-- helping us to identify the
quantization group, even in the solution manifold. This can be illustrated
by means of the action of the would-be semi-invariance generator in the
Abelian case, $\frac{\partial}{\partial \theta_\mu^L}$, on the
symplectic potential (covariantly written in a spatial hypersurface $\Sigma$):

\[
\vartheta_\sigma \equiv \int _{\Sigma}d\sigma_{\mu} \pi^{\mu}_a d \varphi^{a} = \int _{\Sigma}d\sigma_{\mu} \frac{\partial
\mathcal L_\sigma}{\partial \varphi^a_\mu} d\varphi^a = \int _{\Sigma}d\sigma_{\mu}  \theta_a
^{\,L\mu}  \theta^{\,L\,a},
\]
\[L_{\frac{\partial}{\partial
\theta_\mu^{\,L}}}\vartheta_\sigma = \int _{\Sigma}d\sigma_{\mu} \theta^{\,L}
\]

\noindent But $\theta^L\equiv U^{-1}d U$ is no longer a total differential when dealing
with non-Abelian Lie groups; indeed, it verifies $\partial_\mu
\theta^{\,L}_\nu-\partial_\nu
\theta^{\,L}_\mu=[\theta^{\,L}_\mu,\theta^{\,L}_\nu]\not=0$. So, we are not able to
give the complete dynamical symmetry of the system on the spot.

Although the procedure followed above in the finite-dimensional case might be
formally applied, the normal ordering ambiguities eventually appearing are more
involved in field theory, giving rise to problems quite analogous to the non-renormalizability
arising in the usual canonical quantization approach.
However, this symmetry obstruction does disappear if we restrict to NLSM on just
co-adjoint-like orbits of the corresponding group, governed by a \textit{partial trace}
Lagrangian.

\subsubsection{Partial trace.}

Let us replace the target manifold for the fields in (\ref{trazatoa}) (i.e. $G\equiv SU(2)$) by 
a coset $G/G_\lambda$, $G_\lambda=U(1)$ being the isotropy subgroup of a given Lie algebra element 
$\lambda=\lambda^a \tau_a$ under the adjoint action $\lambda \to g \lambda g^{-1}$ of $SU(2)$. To be precise $\lambda$ should have been defined as an element of the dual of the Lie algebra, which is equivalent to the Lie algebra since $SU(2)$ is semisimple.
Now, the (total-trace) NLSM Lagrangian (\ref{trazatoa}) is replaced by 
the partial-trace one:
\be
 \mathcal{L}_{G/G_\lambda} =
\um {\rm Tr}_{G/G_\lambda}(\theta^{\,L}_{\mu} \theta^{\,L \mu})
 \equiv \um {\rm Tr}_{G}([\lambda,\theta^{\,L}_{\mu}] [ \lambda,\theta^{\,L \mu}])\,.
\label{trazapar}
\ee

It can be realized that, defining
\[
\Lambda(x)\equiv U(x) \lambda U(x)^{-1}\equiv \Lambda^a(x)\tau_a ,\quad U(x) \in G\,,
\]
we have an alternative way of writing (\ref{trazapar}) as 
\be
\mathcal{L}_{G/G_\lambda}=\um {\rm Tr}_{G}(\Lambda_{\mu} \Lambda^{\mu})=\um K_{ab} \Lambda^a_{\mu} \Lambda^{b\,\mu}, \label{nesfera}.
\ee 
where $K_{ab}$ is the Killing metric. 
Note that this Lagrangian is singular due to the existence of constraints like, for example, ${\rm Tr}_{G}(\Lambda(x)^2)=
{\rm Tr}_{G}(\lambda^2)\equiv \hbox{constant}$. We shall not deal with constraints 
at this stage. They will be naturally addressed inside our quantization procedure below.

The Lagrangian (\ref{nesfera}) generalizes to field theory that of a point particle constrained to move on a sphere. In \cite{sigmita} it was shown that the Euclidean Group constitutes the addressing symmetry in the case of $\mathbb S^2$, achieving the expected quantum theory.

\subsubsection{Local Euclidean Group.}

The field analogue to the symmetry of a particle moving on the sphere $\mathbb S^2$ proves to be a local version of the 
Euclidean group and its unitary and irreducible representations are intended to account for the quantization of the 
partial-trace NLSM associated with the group $G=SU(2)$ (the extension to any semi-simple Lie group $G$ is essentially 
straightforward). This group will be parameterized by the local $SU(2)$ parameters $\varphi^a(x)$ and the (co-)adjoint 
parameters $\theta^{L\,a}_{\mu}(x)$ (the superscript $L$ will be omitted in the sequel). Then, the vector fields 
$X_{\varphi^a(x)}\,, a=1,2,3$, generate local internal rotations and the vector fields 
$X_{\theta^{L\,a}_{\mu}(x)}\,,\; a=1,2,3\,, \;\mu=0,1,2,3$, generate the  (co-)tangent subgroup. In terms of these 
variables the group law is:

%
 \begin{eqnarray*}
    U''(x) &=& U'(x) U(x) \\
    \theta''_\mu (x)\, &=& U'(x) \; \theta_\mu(x) \; U'^{\dagger}(x) 
    + \theta'_\mu(x) \\
     \zeta'' 
    &=& \zeta' \zeta \exp \left\lbrace i \int_\Sigma d\sigma^\nu\,
    {\rm Tr}\left[ \lambda
    \left(  U'(x) \; \theta_\nu(x) \; U'^{\dagger}(x) 
    - \theta_\nu(x) \right)\right] \right\rbrace, 
 \end{eqnarray*}
\noindent where $U(x)\equiv U(\varphi\,(x))$. Here, all fields are assumed to be 
defined on the Cauchy surface $\Sigma$, so that, the time translation can not be directly implemented, in  
contrast with the Klein-Gordon case. However, we shall construct an 
explicit Hamiltonian operator to account for the time evolution on the quantum states (see below).


We can immediately compute the corresponding right-invariant vector fields:
\bea
\tilde X^{R}_{\varphi^a (x)} &=& X^{R\;(G)}_{\varphi^a(x)}
-\eta_{ab}^{\phantom{ab}c}\theta^b_\mu(x)\frac{\delta\;\;}{\delta\theta^c_\mu(x)}
+ \eta_{ab}^{\phantom{ab}c}\theta^b_\mu(x)\lambda_c n^\mu\Xi\nn\\
\tilde X^{R}_{\theta^a_{\mu}(x)} &=&\frac{\delta\;\;}{\delta \theta^a_{\mu}(x)} \nn\\
\tilde X^{R}_{\zeta} &=& \hbox{Re}(i\zeta \frac{\partial}{\partial\zeta})\equiv \Xi\,,\label{ym-XR}
\eea

\noindent closing the Lie algebra of the local Euclidean Group:
\begin{eqnarray}
\left[\tilde X^{R}_{\varphi^a (x)},\tilde X^{R}_{\varphi^b (y)} \right] 
&=& 
- \eta_{ab}^{\phantom{ab}c} \delta(x-y) \tilde X^{R}_{\varphi^c (x)}
\nn \\
\left[\tilde X^{R}_{\varphi^a (x)}, \tilde X^{R}_{\theta^b_{\mu}(y)}\right] 
&=&
- \eta_{ab}^{\phantom{ab}c} \delta(x-y) \left(\tilde X^{R}_{\theta^c_{\mu}(x)}  -
   \lambda_c n^\mu \; \Xi \right) \label{conmutadores}
\\
\left[\tilde X^{R}_{\theta^a_{\mu}(x)},\tilde X^{R}_{\theta^b_{\nu}(y)} \right]
&=& 
0\;.\nn
\end{eqnarray}
%

We can also compute the  left-invariant vector fields:
\bea
\tilde{X}^{L}_{\varphi(x)}&=&X^{L\;(G)}_{\varphi(x)}\nn\\
\tilde{X}^{L}_{\theta_\mu(x)}&=&U\frac{\!\!\delta}{\delta \theta_\mu(x)}U^\dag-(U^\dag\lambda U-\lambda)n^\mu\Xi\nn\\
\tilde {X}^{L}_{\zeta} &=& \hbox{Re}(i\zeta \frac{\partial}{\partial\zeta})\equiv \Xi\,,\label{sigma-XL}
\eea
closing the same Lie algebra except for opposite structure constants.
Directly from the group law or by duality on (\ref{sigma-XL}) the left-invariant $1$-form in 
the $\zeta$-direction, $\Theta_\sigma$, can be computed 

\be
\Theta_\sigma^{G/G_\lambda}= - \int_\Sigma d{\sigma}^\nu \hbox{Tr}
\left( (\Lambda-\lambda)\delta \theta_\nu \right)+\frac{d\zeta}{i\zeta}\,.    \label{TetaSigma}
\ee
Note that only the time-like component of those $\theta_\mu$ in the coadjoint orbit defined by $\lambda$ will contribute
to the solution manifold. In fact, the characteristic module is generated by:
\[
   \mathcal G_{\Theta_\sigma^{G/G_\lambda}}= 
        \langle \lambda^a  \tilde X^{L}_{\varphi^a (x)},\; 
                \lambda^b n_\mu \tilde X^{L}_{\theta_{\mu}^b(x)}, \;
                \tilde X^{L}_{\theta_{\nu}^c(x)} - n^\nu \tilde X^{L}_{\theta_{\rho}^c(x)} n^\rho
         \rangle.
\]
According to the general formalism, Sec. 3, the Noether invariants are the following:
\bea
I_{\varphi}&=&[\Lambda n_\mu,\,\theta^\mu] \equiv \mathbb L\nn\\
I_{\theta^\mu}&=&(\Lambda-\lambda) n_\mu \,,
\label{NoetherSigma}
\eea
although only $\mathbb L$ and $\mathbb S \equiv n^\mu I_{\theta^\mu}$ are independent, the basic ones, and parameterize
the solution manifold.
In terms of these Noether invariants, the quantization 1-form can be expressed as
\be
\Theta^{G/G_\lambda}_\sigma=\int_\Sigma d\sigma^\nu
\hbox{Tr}\left(\mathbb S \,\delta [\mathbb L,I_{\theta^\nu}]\right) +\frac{d\zeta}{i\zeta}\,.
\ee
The basic Poisson brackets read
\bea 
\left\{\mathbb{L}_a(\vec{x}),\mathbb{L}_b(\vec{y})\right\} &=& -\eta_{ab}^{\phantom{ab}c}
\mathbb{L}_c(\vec{x})\delta(\vec{x}-\vec{y}),\nn\\
 \left\{\mathbb{L}_a(\vec{x}),\mathbb{S}_b(\vec{y})\right\} &=& -\eta_{ab}^{\phantom{ab}c} 
\mathbb{S}_c(\vec{x})\delta(\vec{x}-\vec{y})
 +\eta_{ab}^c\lambda_c\delta(\vec{x}-\vec{y})\,.
\eea 

Even though we have not included the Poincar\'e subgroup, the time evolution (we shall choose
$\Sigma= {\mathbb R}^3$ in the time direction, i.e., $\rm{d}\sigma_\mu\to\rm{d}^3x$)
can be realized on the solution manifold
by giving a Hamiltonian function of the basic Noether invariants. The Hamiltonian, 
\be
\mathbb{H}=\frac{1}{2}\int \rm{d}^3x\{\mathbb{L}^2+(\vec \nabla \mathbb{S})^2\} \,,
\label{Amiltonio}
\ee
reproduces the classical equations of motion in terms of the Poisson brackets above.

In order to achieve the quantization of the present system, we must find a polarization sub-algebra. 
It is given by the characteristic module together with half of the conjugated pairs: 

\[
   \mathcal P = \langle \lambda^a  \tilde X^{L}_{\varphi^a (x)},\;
                 n_\mu \tilde X^{L}_{\theta_{\mu}^b(x)},\;
                \tilde X^{L}_{\theta_{\nu}^c(x)} - n^\nu \tilde X^{L}_{\theta_{\rho}^c(x)} n^\rho
         \rangle.
\]
%

An irreducible representation of the group is given by the action of the right-invariant vector fields of 
the group on the complex functions valued over the group manifold, provided that these functions are polarized 
and satisfy the U(1)-function condition:
 
 \[
    \mathcal P \Psi = 0, \qquad \Xi \Psi = i \Psi.
 \]

It can be easily checked that such functions (now true \textit{wave functions}) are of the form:
\be
\psi=\zeta e^{-i\int_\Sigma \d\sigma^\mu\hbox{Tr}\left(\lambda(U^\dag\theta_\mu U-\theta_\mu)\right)}
\Phi((\Lambda-\lambda))
\label{wavefunction}
\ee
where $\Phi$ is an arbitrary function of its argument. 
  
The action of the right-invariant vector fields preserve the space of polarized wave functions, 
due to the commutativity of the left and right actions
as  already stated, so that it is possible to define an action of them on the arbitrary factor 
$\Phi$ in the wave functions. It is not difficult to 
demonstrate that on this space of functions the basic quantum operators acquire the following expression:
\bea
\hat{S}_a\Phi\equiv i\zeta^{-1} e^{i\int_\Sigma \d\sigma^\mu\hbox{Tr}\left(\lambda(U^\dag\theta_\mu U-\theta_\mu)\right)}
\,n_\nu \tilde{X}^R_{\theta^a_\nu}\psi&=&(\mathbb{S}_a- \lambda_a)\Phi\nn\\
\hat{L}_a\Phi\equiv-i\,\zeta^{-1} e^{i\int_\Sigma \d\sigma^\mu\hbox{Tr}\left(\lambda(U^\dag\theta_\mu U-\theta_\mu)\right)}
\tilde{X}^R_{\varphi^a}\psi&=&
\eta^{\phantom{ab}c}_{ab} \;\mathbb{S}^b \frac{\!\!\delta}{\delta \mathbb{S}^c} \Phi\,,
\eea
where the operator $n_\mu \tilde X^{R}_{\theta^a_{\mu} (x)}$ can be redefined with 
the addition of $\lambda$ to fix the vacuum expectation value to zero. 

On the quantum representation space we can construct the Hamiltonian operator 
\be
\hat{H}\Phi=\frac{1}{2}\int \d^3x\{\hat{L}^2 + (\vec \nabla \hat{S})^2\}\Phi\,,
\ee
that  represents the classical Hamiltonian $\mathbb H$ without ordering ambiguity, due to its quadratic 
expression in terms of the basic operators.
It must be emphasized that this operator preserves the Hilbert space of quantum states.

Thus, it becomes evident at this point that the domain of the wave functions has been naturally selected 
without imposing any constraint condition as such. Our quantization program chooses as the basic observables
the $\tilde X^{R}_{\varphi (x)}$ operators, playing the role of ``generators of translations'' in the internal 
parameter space, and the conjugated ones $n_\mu (x) \tilde X^{R}_{\theta_{\mu} (x)}$, playing that of ``field'' 
operators. Note that had we resorted to canonical quantization we should have postulated basic commutators of 
the generic form $[\varphi(x),\frac{\partial}{\partial \varphi(y)}]= \delta(x-y)$, in deep contrast with 
(\ref{conmutadores}).

Let us remark that even though we have discarded the Poincar\'e symmetry as a subgroup of our quantization group $\tilde{G}$, 
the extended local Euclidean group, and in particular the time translation, we have provided a quantum Hamiltonian preserving 
the Hilbert space of the quantum representation. This means that we can proceed with any computation involving the time evolution 
without disturbing the physical system as a whole. Namely, we can construct a perturbative theory in the 
``Heisenberg picture''\cite{Landau}, that is to say, evolving the wave functions, originally defined on the Cauchy hypersurface, 
with the exponential of the total Hamiltonian. The explicit perturbation series can be achieved according to the Magnus 
expansion \cite{Magnus} which guarantees unitarity to each order. Also, if the actual Hamiltonian can be decomposed into two pieces
each one preserving the original Hilbert space, one of them considered as a free Hamiltonian $\hat{H}_0$, the other as an 
interaction Hamiltonian, $\hat{H}_{int}$, it is then possible to achieve the perturbation theory in two steps. In that case, 
in the first step we arrive at some sort of ``free'' theory to which we apply the perturbation addressed by $\hat{H}_{int}$, in a way 
that realizes the more standard Dyson expansion in the ``interaction picture''. In Ref. \cite{rotonescoloraos} we analyzed this 
scheme and discussed the differences with ordinary perturbation approach, in particular the fact that we start from a Hamiltonian 
$\hat{H}_0$ whose classical theory possesses the same topology as that of the total Hamiltonian, a relevant ingredient to achieve 
a unitary, renormalized theory.


%
%
%
%

{

This group-quantization of non-linear sigma models is being further developed in relation with relevant physical applications 
in alternatives to the Higgs-Kibble mechanism of mass generation \cite{mym_group} (see also \cite{MPLA,maneeds}). 
%

\section*{Acknowledgments}
Work partially supported by the Fundaci\'on S\'eneca,
Spanish MICINN and Junta de Andaluc\'\i a under projects
08814/PI/08, FIS2008-06078-C03-01 and FQM219-FQM1951, respectively. F.F.
L\'opez-Ruiz thanks C.S.I.C. for an I3P grant. M. Calixto thanks the
Spanish MICINN for a mobility grant PR2008-0218. All the 
authors wish to thank  J.L. Jaramillo and E. 
S\'anchez-Sastre for useful discussions.


\end{document}